\documentclass[12pt]{article}
\usepackage{amssymb}
\def\beq{\begin{equation}}
\def\eeq{\end{equation}}
\usepackage[all,2cell,dvips]{xy}
\newtheorem{proposicion}{Proposition}

\def\IR{\relax{\rm I\kern -.18em R}}

\begin{document}
\title{Non-perturbative analysis of the gravitational energy in Ho\v{r}ava Theory }
\author{ \Large J. Bellorin*, A. Restuccia**, A. Sotomayor***}
\maketitle{\centerline{*Departamento de F\'{\i}sica}}
\maketitle{\centerline{Universidad Sim\'on Bol\'{\i}var }
\maketitle{\centerline{**Departamento de F\'{\i}sica}}
\maketitle{\centerline{Universidad de Antofagasta}
\maketitle{\centerline{**Departamento de F\'{\i}sica}}
\maketitle{\centerline{Universidad Sim\'on Bol\'{\i}var }
\maketitle{\centerline{***Departamento de Matem\'aticas}
\maketitle{\centerline{Universidad de Antofagasta}}
\maketitle{\centerline{e-mail: jorgebellorin@usb.ve,
arestu@usb.ve, asotomayor@uantof.cl }}

\begin{abstract}We perform a non-perturbative analysis of the constraints of the
Ho\v{r}ava Gravitational theory. In distinction to Einstein
gravity the theory has constraints of the first class together
with second class ones.  We analyze the consequences of having to
impose second classes constraints at any time in the quantum
formulation of the theory. The second class constraints are
formulated as strongly elliptic partial differential equations
allowing a global analysis on the existence and uniqueness of the
solution. We discuss the possibility of formulating the theory in
terms of a master action with first class constraints only. In
this case the Ho\v{r}ava theory would correspond to a gauged fixed
version of the master theory. Finally we obtain , using the
non-perturbative solution of the constraints, the explicit
expression of the gravitational energy. It is, under some
assumptions, always positive and the solution of Ho\v{r}ava field
equations at minimal energy is the Minkowski metric.
\end{abstract}

\section{Introduction}Recently, Ho\v{r}ava \cite{Horava} proposed a gravitational theory
with Lifshitz-like anisotropic scaling at short distances: \[
t\rightarrow b^zt,\hspace{3mm}x\rightarrow bx. \]

The formulation breaks the relativistic symmetry at short
distances with the idea of regaining it at large distances. The
benefit would be to obtain a power counting renormalizable theory
of gravity. The anisotropic scaling between time and space allows
to include in the action high enough spatial derivatives which
contribute to the interactions and to the propagators improving
the $UV$ properties of the theory. In order to obtain a
renormalizable theory $z$ should be equal to the number of
spatial dimension and all terms compatible with the gauge
symmetry should be included in the action.

In distinction to General Relativity, Ho\v{r}ava Gravity is
restricted not only by first class contraints but also by second
class constraints which are potentially dangerous because they may
introduce non-localities in the quantum formulation of the
theory. Directly related to this analysis and an important aspect
of any gravitational theory is to determine the gravitational
energy and to established its positivity. The positive mass
theorem plays a fundamental role in General Relativity. It has
firstly proved in \cite{Deser,Schoen,Witten} that for
asymptotically flat space-times the total energy-momentum in
General Relativity is well defined, it is greater or equal to
zero and it vanishes for flat space-time. We will discuss the
existence and uniqueness of the solution to the constraints and
will prove the positivity of the gravitational energy for the
Ho\v{r}ava theory in the large distance regime. Moreover the
expression for the gravitational energy we will obtain remains
valid even when one includes all the interacting terms
corresponding to the $z=3$ Ho\v{r}ava theory.

\section{The Ho\v{r}ava action}
Ho\v{r}ava theory is formulated on a foliated manifold $
\mathcal{M}=\Sigma\times \mathbb{R}$ where $\Sigma$ is a three
dimensional Riemann manifold which we will assume to be complete,
connected and asymptotically flat.

The theory is expressed in terms of an $ADM$ formulation where
the metric is given by \beq
ds^2=\left(-N^2+N_iN^i\right)dt^2+2N_idx^idt+g_{ij}dx^idx^j.\eeq

The extrinsic curvature of $\Sigma$, denoted $K_{ij}$, satisfies
\beq K_{ij}=\frac{1}{2N}\left(\dot{g}_{ij}-\nabla_iN_j\right)\eeq
where the lapse $N$ is assumed to be different from zero.

The Hamiltonian of Ho\v{r}ava theory is generically given by \beq
\int
dtd^3x\sqrt{g}N\left(G^{ijkl}K_{ij}K_{kl}+V\left(g,N\right)\right)
\eeq where \beq
G^{ijkl}=\frac{1}{2}\left(g^{ik}g^{jl}+g^{il}g^{jk}-\lambda
g^{ij}g^{kl}\right), \eeq $\lambda$ is a dimensionless parameter
which may be included because each term is separately a tensor
under spatial diffeomorphism the gauge symmetry of the theory.
More precisely the above action is invariant under diffeomorphism
on $\Sigma$ and reparametrization on $t$: \beq
\widetilde{t}=f(t),\hspace{3mm}\widetilde{x}=g(x).\eeq The
symmetry (5) may be enlarged by considering anisotropical
conformal gauge transformations, which are relevant in the
formulation of conformal gravity [Ho\v{r}ava, 2012]\cite{Horava1}.
We will only be concerned with transformations (5).

In the potential $V(g,N)$ one should include all possible local,
up to $z=3$ terms, compatible with the gauge symmetries of the
theory.

Typical $z=3$ terms are $\nabla_kR_{ij}\nabla^kR^{ij},R\Delta
R,R^{ij}\Delta R_{ij}$ which contribute to the interactions but
also modify the propagators improving the $UV$ behaviour of the
theory. Other $z=3$ terms such as $R^3,RR_{ij}R^{ij}$ are pure
interacting terms.
\section{The hamiltonian in the large distance regime}

The most general Hamiltonian describing the large distance regime
in Ho\v{r}ava theory is given by \cite{Donelly,Bellorin1}

\beq H=\int d^3x\left(N\mathcal{H}+N_i\mathcal{H}^i+\sigma
\phi\right)+E_{ADM}-2\alpha \phi_N \eeq

\beq \mathcal{H}\equiv
G_{ijkl}\frac{\pi^{ij}\pi^{kl}}{\sqrt{g}}+\sqrt{g}\left(-R+\alpha\left(2\nabla_ia^i+a_ia^i\right)\right)\eeq

\beq \mathcal{H}^i \equiv -2\nabla_j\pi^{ji}+\phi\partial^iN \eeq

\beq \phi_N\equiv \oint d\Sigma_i\partial_iN, \eeq where we have
assumed that $\lambda\neq\frac{1}{3}.$ For $\lambda=\frac{1}{3},
G^{ijkl}$ is not invertible. The theory with $\alpha=0$ for any
value of $\lambda$ is equivalent to General Relativity
\cite{Bellorin2,Bellorin3}. The terms which depends on $N$ and
their derivatives in the potential were introduced in \cite{Blas}.

As in General Relativity the $ADM$ energy in included as the
boundary term. This is necessary in order to obtain the equation
of motion under variations $\delta g_{ij}$ asymptotically of
order $ O(r^{-1})$ \cite{Regge}. Similarly, the flux of $N$ at
spatial infinity, $\phi_N$, cancels a non zero contribution
coming from $\delta(2\alpha N\nabla_ia^i)$ for variations of $N$
with asymptotic decay $\delta N=O(r^{-1}).$

$ \mathcal{H}^i=0$ is a first class constraint related to the
generator of spatial diffeomorphisms. $ \mathcal{H}=0$ is a
second class constraint. In order to preserve it we obtain a
$PDE$ for the Lagrange multiplier $\sigma,$

\beq
\partial_i\left(N\sqrt{g}g^{ij}\partial_j\left(\sigma/
N\right)\right)=\partial_i\left(-\gamma
N^2\nabla^i\pi+N^2G_{klm}^i\pi^{lm}\right). \eeq

The Dirac procedure ends at this stage.

\section{The non-perturbative analysis of the constraints}

In order to analize the constraints $\mathcal{H}=0$ and the
equation for the Lagrange multiplier we consider the conformal
transformation \beq\begin{array}{l}g_{ij}=e^\varphi
\widehat{g}_{ij}
\\ \pi^{ij}e^{-\varphi}\widehat{\pi}^{ij}
\end{array}\eeq where the Riemannian metric $ \widehat{g}_{ij}$
satisfies $ \det \widehat{g}_{ij}=1$.

We also consider the decomposition of
$\pi^{ij}=\widehat{\pi}^{ij}_T+\frac{1}{3}\widehat{g}^{kj}\widehat{\pi}$
where $\widehat{g}^{ij}$ is the inverse of $\widehat{g}_{ij}$ and
$\widehat{\pi}^{ij}_T$ is traceless:
$\widehat{g}_{ij}\widehat{\pi}^{ij}=0.$

The transformation $\left(g_{ij},\pi^j\right)\rightarrow
\left(\widehat{g}_{ij},\widehat{\pi}^{ij}_T,\varphi,\widehat{\pi}\right)$
is canonical: \beq \int_\Sigma
d^3x\pi^{ij}\dot{g}_{ij}=\int_\Sigma
d^3x\left(\widehat{\pi}^{ij}_T\dot{\widehat{g}_{ij}}+\widehat{\pi}\dot{\varphi}\right).
\eeq

In the new variables the constraint $ \mathcal{H}=0$ becomes
\beq\partial_i\left(e^{\beta\varphi}\widehat{g}^{ij}\partial_je^{\frac{\xi}{2\alpha}}\right)+Ge^{\frac{\xi}{2\alpha}}=0\eeq
where \beq
G=-\frac{e^{\beta\varphi}}{4\alpha}\left[e^{-2\varphi}\left(\widehat{\pi}^{ij}_T\widehat{\pi}_{Tij}+{\left(3-9\lambda\right)}^{-1}\right)\widehat{\pi}^2-
\widehat{R}-\beta\partial_i\varphi\partial^i\varphi \right]\eeq
and $\xi\equiv \alpha\ln |N|+\varphi.$

We notice that $G$ depends on the canonical pairs $\left(
\widehat{g}_{ij},\widehat{\pi}^{ij}_T\right)$ and
$\left(\varphi,\widehat{\pi}\right).$ It does not depend on $N$.
 Equation (13) is a linear partial differential equation on
 $e^{\frac{\xi}{2\alpha}}$. It is strongly elliptic. It is
 convenient to consider $e^{\frac{\xi}{2\alpha}}=1+u$ and to
 define a suitable space of functions for $u$.

 We can prove the following proposition \cite{Bellorin4}:

 \begin{proposicion} Given an asymptotically flat Riemannian manifold
 with $C^\infty$ metric $g_{ij}$ and momenta $\pi^{ij}$, with the
 asymptotic behaviour
 \begin{eqnarray*}& & g_{ij}=\delta_{ij}+O(r^{-1}) \\ && \pi^{ij}=O(r^{-2}), \end{eqnarray*}
 then the asymptotic solution for (13) exists, it is $C^\infty$
 and the asymptotic behaviour is \[u=O(r^{-1}).\] \end{proposicion}

 We are then motivated to introduce the space $\widehat{C}^1$:

 \[\widehat{C}_1(\Sigma)=\left\{ u\in C^1(\Sigma):u=O(r^{-1})\mathrm{\:when\:}r\rightarrow\infty \right\}.\]

 For $u$ and $v$ in $ \widehat{C}_1(\Sigma)$ we define the
 bilinear functional
 \[\left(u,v\right)=\int_\Sigma d^3x\left(e^{\beta\varphi}\widehat{g}^{ij}\partial_iu\partial_jv+Guv\right).\]
 If $ G\geq0$ on $ \Sigma$ then $(u,v)$ defines an internal
 product in $\widehat{C}_1(\Sigma).$ We denote $
 \widehat{\mathcal{H}}^1$ the Hilbert space obtained by the
 completion of $ \widehat{C}_1(\Sigma)$ with respect to the norm
 induced by the above internal product.

 The assumption $G\geq0$ will be essential to prove existence and
 uniqueness for the solution of the constraint. If $G$ is not
 positive the operator \[\mathcal{O}\equiv -\partial_i\left(e^{\beta\varphi}\widehat{g}^{ij}\partial_j\cdot\right)+G\cdot\]
 may have a nontrivial kernel $K$. In that case the solution $u$
 to the constraint \beq \mathcal{O}u=-G\eeq exists if and only if
 $G$ is orthogonal to $K$. Even if this condition is satisfied the
 solution would not be unique.

 We can prove the following propositions \cite{Bellorin4}.

 \begin{proposicion}Given a complete, connected, asymptotically flat Riemannian
 manifold $ \Sigma$ and momenta $ \pi^{ij}$ satisfying $G\geq0$,
 there always exist in $ \widehat{\mathcal{H}}^1$ a unique weak
 solution to the constraint (15).\end{proposicion}

\begin{proposicion}
 Under the previous assumptions and considering a $C^\infty$
 metric $g_{ij}$ and momenta $\pi^{ij}$, the weak solution is
 $C^\infty$.\end{proposicion}

\begin{proposicion}
 Under the previous assumptions the solution satisfies $1+u\geq0.$
 Hence we may identify \[e^{\frac{\xi}{2\alpha}}\equiv 1+u\geq 0.\]\end{proposicion}

\begin{proposicion}
 Under the previous assumptions the solution for the Lagrange
 multiplier $ \sigma$ exists and is unique. The asymptotic
 behaviour is \[\sigma=O(r^{-2})\] when $r\rightarrow\infty$.\end{proposicion}

 These propositions prove under above assumptions, the existence and
 uniqueness of the solution to the constraint $ \mathcal{H}=0$ and
 of the equation for the Lagrange multiplier.

 We can finally obtain the following results concerning the
 gravitational energy of the Ho\v{r}ava gravity.

 The gravitational energy is given by the flux of the $ \xi$
 field:
 \[E=-2\oint d\Sigma_ig^{ij}\partial_j\xi.\]

 Under the previous assumptions and for $\alpha\leq0$ the
 gravitational energy $E$ is positive.

 The energy $E$ has a minimum for $\xi=0$. At the minimum and
 using the field equations for Ho\v{r}ava gravitational theory we
 obtain
 \beq\begin{array}{c} g_{ij}=\delta_{ij},\pi^{ij}=0 \\ N=1   \end{array}
 \eeq in the gauge $N^i=0$.

 It is important to distinguish this space-time metric from one
 obtained by performing an anisotropic conformal transformation on
 a Lifchitz space-time. If fact, the field equations for the
 action we are considering are invariant under an isotropic
 conformal transformation (exactly the same one that leave
 invariant Einstein equations for General Relativity). The metrics
 are equal but the symmetries of the underlying spaces are different.

 \section{Conclusions}We presented a non-perturvatibe analysis of
 the existence and uniqueness of the second class contraints of
 the Ho\v{r}ava Gravity in the large distance regime where $z=3$
 interacting terms and their contributions to the propagator are
 not relevant. The action we considered includes all the $z=1$
 possible contributions (without the cosmological term). We
 provide a sufficient condition ensuring the existence and
 uniqueness of the solution to the constraints. After the
 elimination of : a) the lapse $N$, from the solution for the
 $\xi$ field, and its conjugate momentum from the second class
 contraints, b) the gauge field $N^i$ and the longitudinal part of
 $\pi^{ij}$ from the first class constraint, we are then left with
 the physical degrees of freedom in terms of the conjugate pairs
 $\left(\pi_T^{ij},\hat{g}_{ij}\right)$ and
 $\left(\hat{\pi},\varphi\right)$. There is one additional degree
 of freedom with respect to General Relativity. The interesting
 point of our presentation is that the gravitational energy of the
 theory is directly expressed as the flux of the $\xi$ field whose
 solution is obtained directly from the second class constraint $
 \mathcal{H}=0$. We gave a sufficient condition ensuring the
 positivity of the gravitational energy for Ho\v{r}ava theory.

 Although the second class contraints can be qualitatively solved
 and the solution renders a positive gravitational energy still
 the goal of a renormalizable field theory has not been reached.
 The reason is that the complicated (when $z=3$ contributions are
 included) second class constraints have to be imposed at any time
 and the possibility to explicitly solve them without introducing
 non-localities is far for been obtained. An alternative approach
 to handle this problem arises from the structure of the metric of
 Poisson bracket of the second class constraint. Their
 contribution to the measure of the path integral is through the
 square root of its determinant. In general, even with the
 inclusion of $z=3$ terms in the hamiltonian, the square root
 reduces only to the bracket $\left\{\mathcal{H},\phi\right\}$.
 This contribution is similar to the one of a theory with first
 class constraint only. In that case the bracket is evaluated
 between the first class constraint and its gauge fixing
 condition. This suggest the existence of a master action with
 first class constraints only from which the Ho\v{r}ava theory
 would arise as a gauge fixed of the master theory. In that case a
 BRST quantization would then solve the problem.

 $\bigskip$

\textbf{Acknowledgments}

A. R. and A. S. are partially supported by Project Fondecyt
1121103, Chile.

\end{document}